\documentclass[%
prb,twocolumn,superscriptaddress,amsmath,amssymb,aps,]{revtex4-1}
\pdfoutput=1

\usepackage{siunitx}
\usepackage{notes2bib}
\usepackage[utf8]{inputenc}
\usepackage[T1]{fontenc}
\usepackage{mathtools}
\usepackage{subfig}
\usepackage{graphicx}
\usepackage{dcolumn}
\usepackage{bm}
\usepackage{verbatim}
\usepackage[normalem]{ulem}
\usepackage[dvipsnames]{xcolor}

\usepackage{hyperref}
\usepackage{multirow}

\usepackage{ragged2e}
\usepackage{caption}
\begin{document}

\author{Susanne Kunzmann}
\affiliation{Experimental Physics, Bielefeld University, Universit\"atsstr 25, 33615 Bielefeld, Germany}
\affiliation{Research Center Future Energy Materials and Systems (RC FEMS), University of Duisburg-Essen,  Forsthausweg 2, 47057 Duisburg, Germany}
\affiliation{Interdisciplinary Centre for Advanced Materials Simulation (ICAMS), Ruhr-University Bochum, Universit\"atsstr 150, 44801 Bochum, Germany}
\author{Thomas Hammerschmidt}
\affiliation{Interdisciplinary Centre for Advanced Materials Simulation (ICAMS), Ruhr-University Bochum, Universit\"atsstr 150, 44801 Bochum, Germany}
\author{Gabi Schierning}
\affiliation{Experimental Physics, Bielefeld University, Universit\"atsstr 25, 33615 Bielefeld, Germany}
\affiliation{Research Center Future Energy Materials and Systems (RC FEMS), University of Duisburg-Essen,  Forsthausweg 2, 47057 Duisburg, Germany}
\affiliation{Center for Nanointegration Duisburg-Essen (CENIDE), University of Duisburg-Essen, Forsthausweg 2, 47057 Duisburg, Germany}
\author{Anna Gr\"unebohm}
\affiliation{Interdisciplinary Centre for Advanced Materials Simulation (ICAMS), Ruhr-University Bochum, Universit\"atsstr 150, 44801 Bochum, Germany}
\affiliation{ Center for Interface-Dominated High Performance Materials (ZGH), Ruhr-University Bochum, Universit\"atsstr 150, 44801 Bochum, Germany}

  \title{Ab initio study of transition paths between (meta)stable phases of Nb and Ta-substituted Nb 
   }
  \date{\today}
\begin{abstract}
Although Niobium is a well characterized material it still shows some anomalies that are not yet understood. Therefore we revisit its metastable phases using density functional theory. First, we systematically compare energies and ground state volumes of chosen crystal structures and discuss possible transition paths to the bcc ground state structure and the energy landscape for tetragonal distortions. Furthermore, we discuss their stability by means of their phonon spectra and vibronic free energies.
Second we analyze the impact of tantalum impurities on phase stability. 
Surprisingly we find new aspects of the energy landscape of the material which have been overlooked so far: A new local energy minimum on the bcc to omega transition path, a flat energy landscape with respect to uniaxial strain along [111] and a considerable stabilization of the $\sigma$ phase by Ta substitution.

\end{abstract}

\pacs{77.80.bg, 77.80.Dj}

\maketitle

%%%%%%%%%%%%%%%%%%%%%%%%%%%%%%%%%%%%%%%%%%%%%%%%%%%%%%%%%%%%%%%
\section{Introduction}

Niobium (Nb) is one of the best studied elemental metals. Nevertheless, and surprisingly, several properties of this material remain unexplained. For instance, it has the highest superconducting transition temperature of all elements at normal pressure,\cite{sauls_2023_effects} but there is still debate about the character of its superconductivity.\cite{SELLERS, struzhkin_superconducting_1997}
Nb is one of the few transition metals that exist in a bcc ground state. However, depending on the boundary conditions, there are many different metastable phases, especially under high pressure \cite{li_shear_2021, errandonea_experimental_2020} or in nanostructures, \cite{wang_consecutive_2018,chattopadhyay_polymorphic_2001} but their relative stability and potential transition pathways are not yet fully understood. 
Remarkably, Bollinger et al.\ \cite{bollinger_observation_2011} experimentally found a change of the slope in the linear thermal expansion coefficients of the bcc state by high-resolution calorimetry at 208~K and related this to a potential martensitic phase transition. However, the signatures of this phase transition were smaller than the detection limit of x-ray diffraction, and hence the nature of this phase transition remained poorly understood.

Martensitic phase transitions are only possible if there is a diffusionless transition path with a moderate energy barrier connecting both the ground state structure and the metastable phase.
Especially in metals, such phase transitions with their complex interplay of phononic, electronic and microstructural properties have been a rich source of research for decades,\cite{grunebohm_unifying_2023} but the question of the driving force of such a transition has not yet been fully resolved.
In Nb, however, some of the typical precursors that usually occur at martensitic phase transitions were evidenced. This includes the occurrence of Kohn anomalies,\cite{nakagawa_lattice_1963, de_gironcoli_lattice_1995, landa_kohn_2018, tidholm_temperature_2020,Aynajian_2008} anomalies in the elastic constants for different pressure ranges, \cite{landa_kohn_2018} and Fermi surface nesting, producing a Van Hove singularity in the electronic density of the states closed to the Fermi level.\cite{liu_first_2011} 
Potential metastable states that may be involved in martensitic phase transformations have been investigated for Nb in several theoretical studies, as this element is often used as prototype material to test simulation methods.\cite{de_gironcoli_lattice_1995,mehl_applications_1996,  fellinger_force-matched_2010,hammerschmidt_topologically_2013} 

The best studied metastable phase of Nb is fcc. \cite{sasaki_fcc_1988,chattopadhyay_polymorphic_2001, wang_consecutive_2018}
Other experimentally found metastable phases include hexagonal-$\omega$ (C32) \cite{li_shear_2021} and Pnma.\cite{errandonea_experimental_2020}  The Pnma phase has been found in experiments.\cite{errandonea_experimental_2020} 
Moreover, the $\omega$ structure is observed in Nb under high pressure at temperatures around 77 K\cite{li_shear_2021} and in thin films.\cite{lee_stress-induced_2022}
These phases have been characterized by density functional theory (DFT)\cite{errandonea_experimental_2020, li_shear_2021} and, in addition, a metastable $\omega$-like structure with vacancies.\cite{lee_stress-induced_2022}
Further studies in literature used DFT to calculate the energies of bcc, hcp and several topologically complex phases (TCP) such as A15, Laves phases and $\sigma$ structures.\cite{hammerschmidt_topologically_2013}
In particular, the A15 phase in Nb-based intermetallic phases, such as Nb$_3$Sn, is famous for the occurrence of both martensitic phase transitions,\cite{godeke_topical_2006} the associated instabilities,\cite{Sadigh_1998_structural} and the occurrence of superconductivity. It is therefore worth examining this TCP phase in elemental Nb in more detail.

Diffusionless transition paths of martensitic transformations among these phases have been studied, particularly the transition from bcc to fcc (Bain path),\cite{schoenecker_theoretical_2011,nnolim_theoretical_2003,craievich_local_1994} and from bcc to hcp.\cite{craievich_structural_1997} 
Complex indirect transition paths have been reported from hcp and bcc to intermetallic phases, such as Laves phases,\cite{natarajan_connecting_2018, KOLLI} and by means of kinetic Monte Carlo simulations from bcc to A15.\cite{xiao_solid-state_2014} For Nb, however, other metastable phases are lower in energy and the possible transition paths are yet unknown.

A peculiarity of Nb experimental works is that Nb crystals are usually contaminated with Ta. The reason is that Nb and Ta have common natural occurrences and that their separation is costly. Both elements are chemically similar, including their crystal structure.\cite{NETE} Similar metastable phases occur in Nb and Ta, e.g., Pnma.\cite{yao_stable_2013} Such additions of Ta have been found to lower the formation energy of metastable TCP phases of Nb~\cite{hammerschmidt_topologically_2013} while the influence on other known metastable phases of Nb is still unknown.

Within this work, we systematically revisit and compare the energies of metastable phases and the details of the diffusionless transition paths connecting these to bcc, both for Nb and the solid solution of Nb$_{(1-x)}$Ta$_x$. We find that Pnma, A15 and $\sigma$ phase are lower in energy than the bct phase and also the energy barrier for the bcc to Pnma transition is \SI{0.2}{eV} lower than the transition barrier from bct to fcc.
Furthermore, the bcc to $\omega$ path shows an additional local energy minimum if extrapolated to larger values of strain. This configuration turns out to correspond to an easy deformation, but it is not stable against the relaxation to a bcc structure strained along [111].

\section{Method}
\begin{figure}[t]
    \subfloat[][]{\includegraphics[clip,trim=12cm 8cm 8.5cm 4.5cm,width=0.15\textwidth]{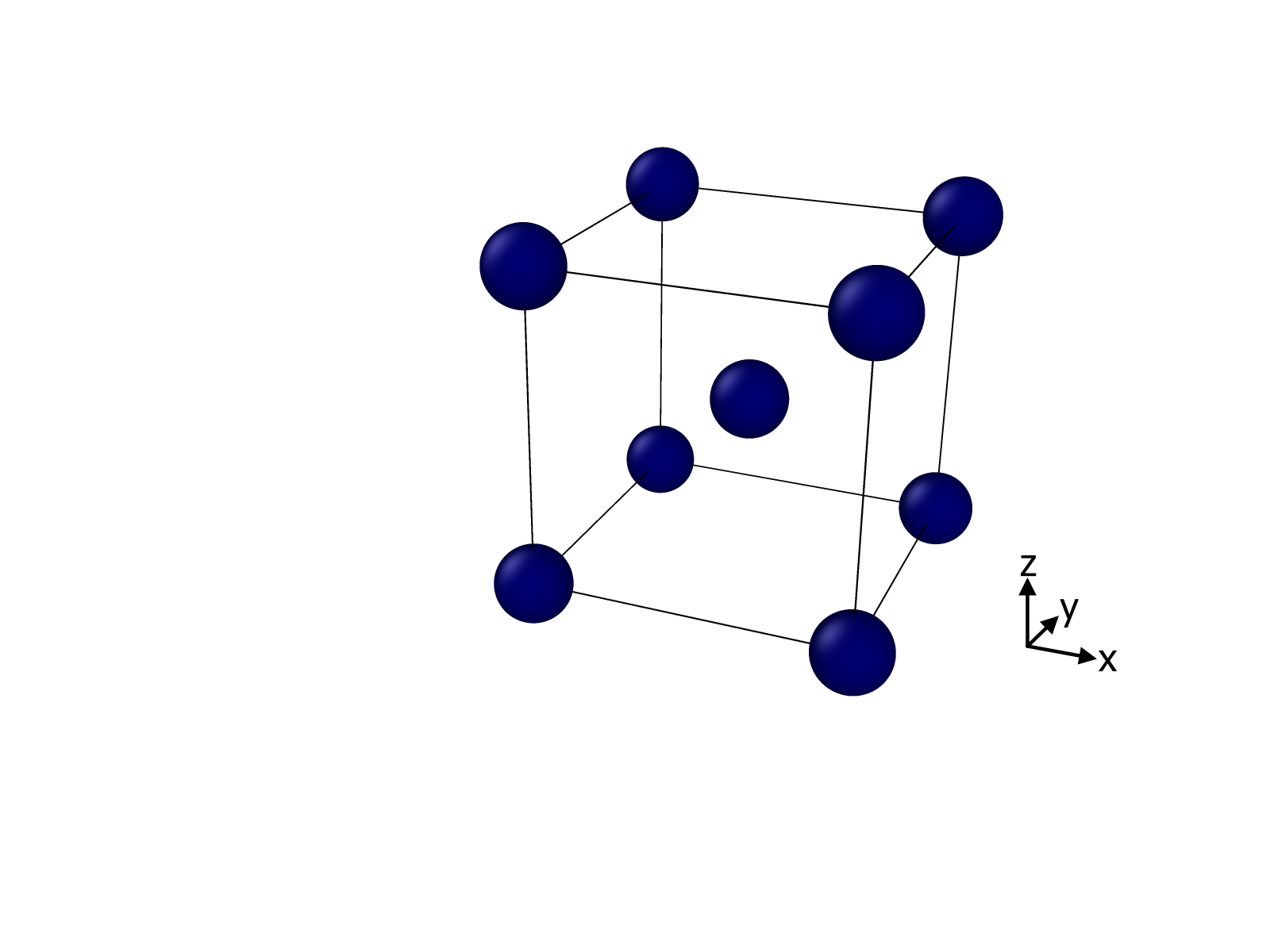}}
    \centering
    \subfloat[][]{\includegraphics[width=0.3\textwidth]{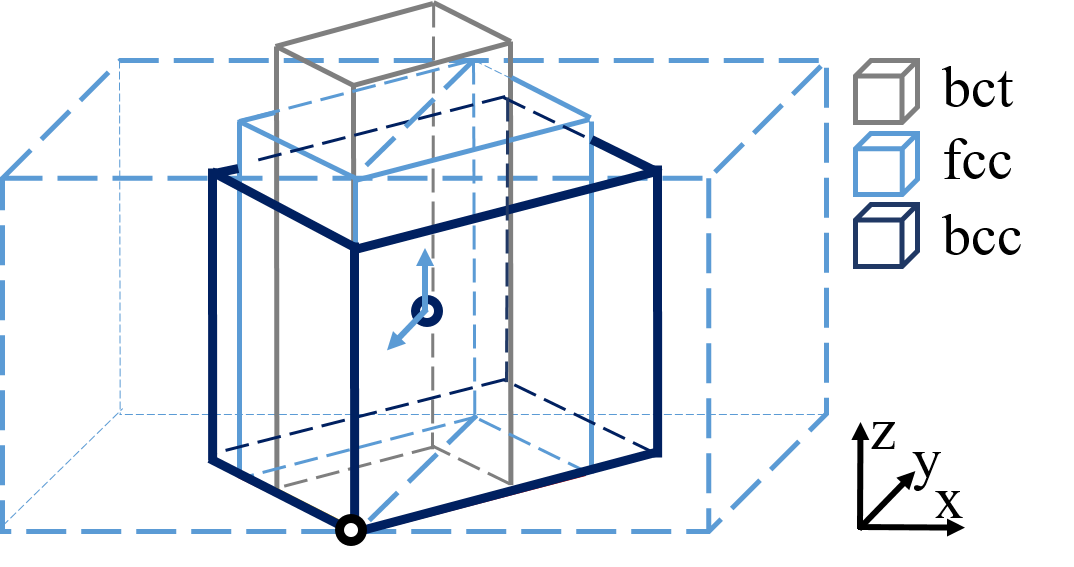}}
    \qquad
    \subfloat[][]{\includegraphics[width=0.3\textwidth]{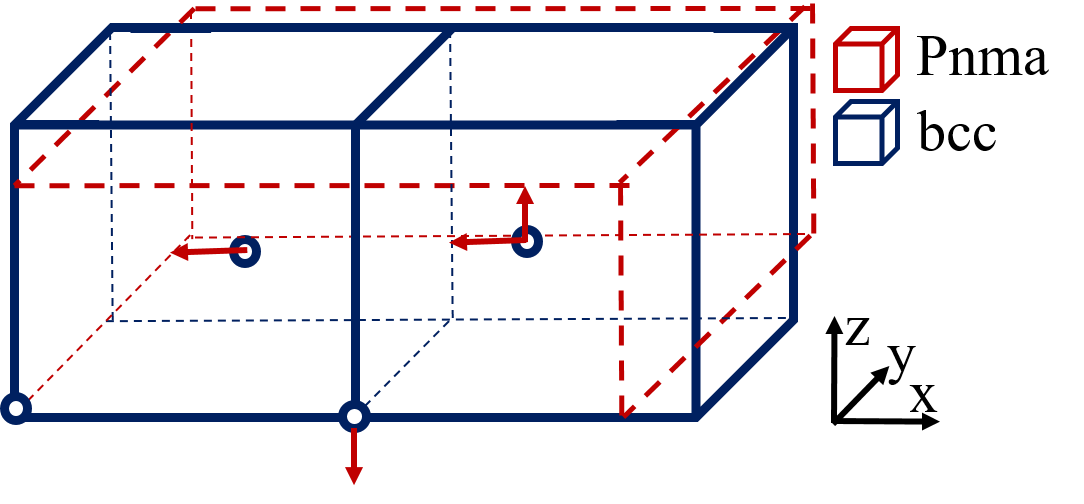}}
    \centering
    \subfloat[][]{\includegraphics[clip,trim=12.5cm 8cm 6.5cm 6cm,width=0.17\textwidth]{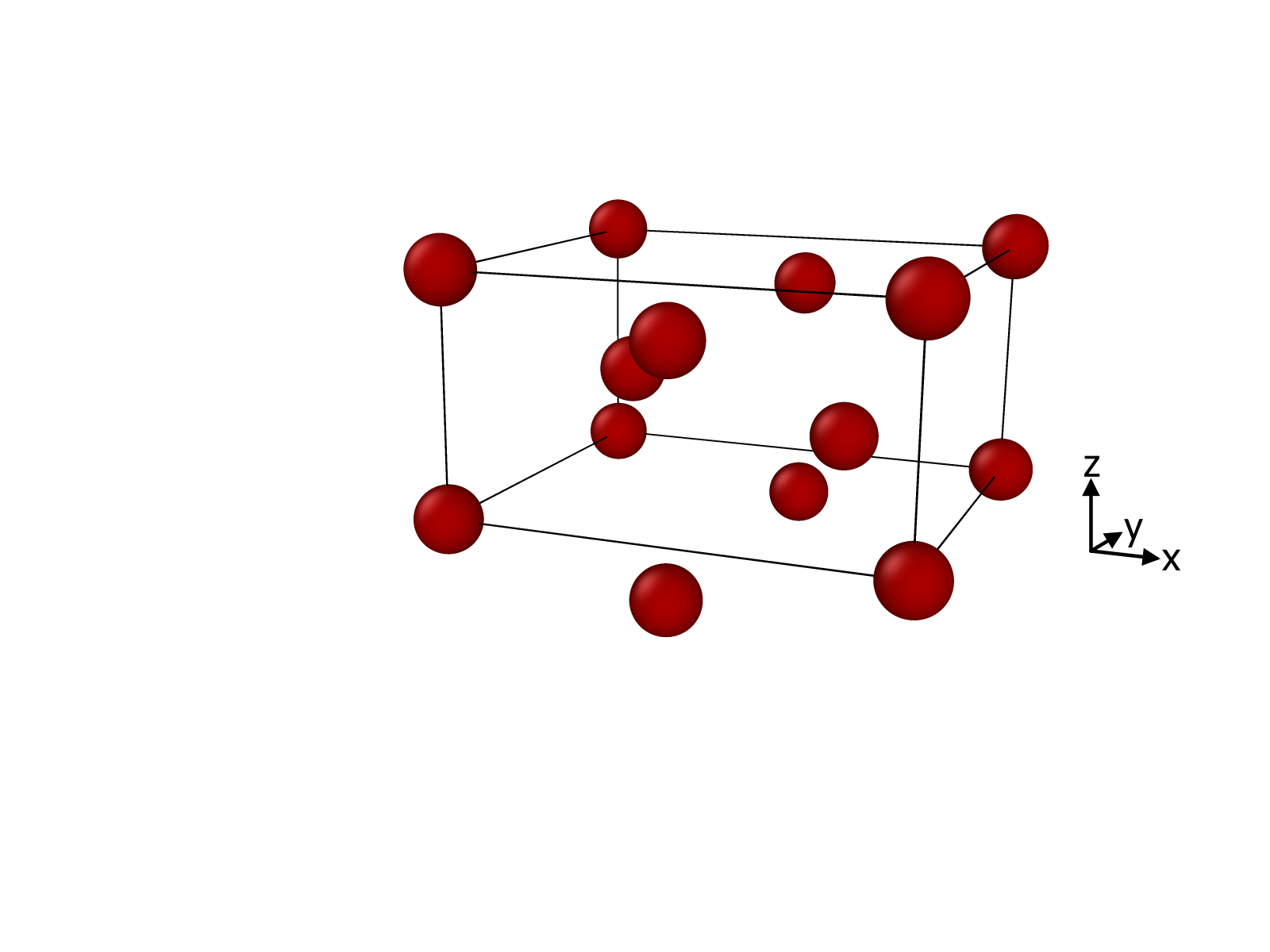}}
    \qquad
    \subfloat[][]{\includegraphics[width=0.3\textwidth]{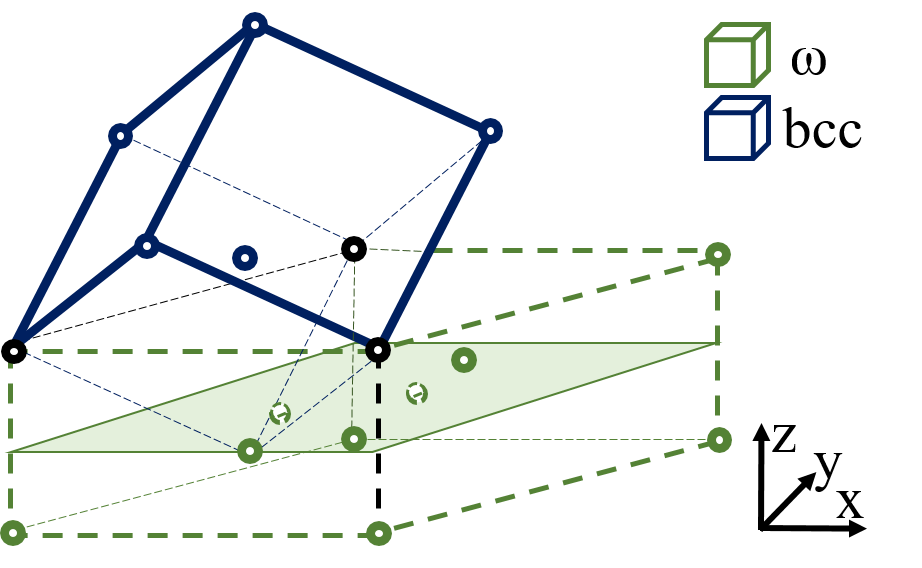}}
    \centering
    \subfloat[][]{\includegraphics[clip,trim=12.5cm 8cm 10cm 6cm,width=0.17\textwidth]{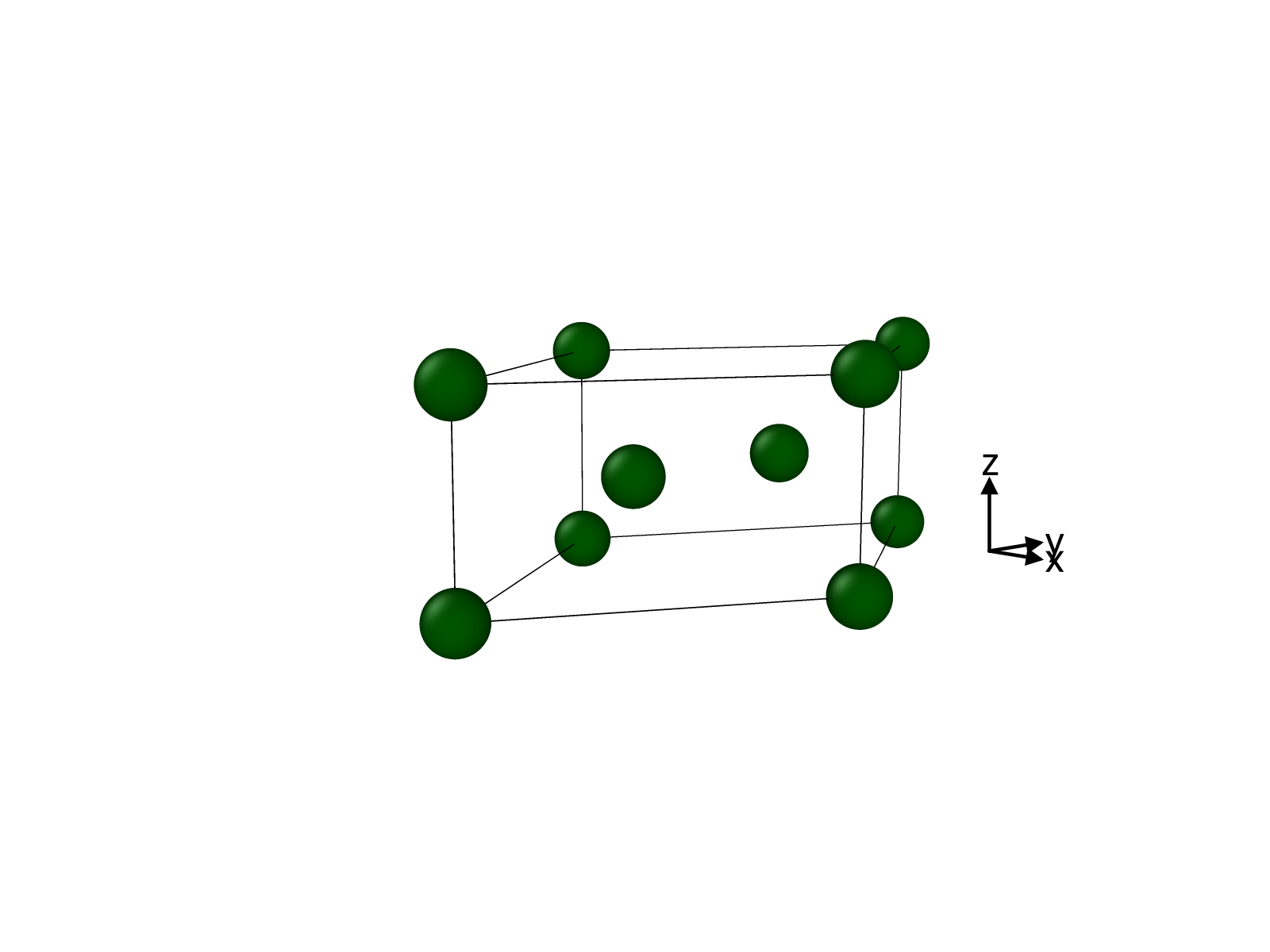}}
    \caption{Schematic representation of (a) the unit cell of the  
    bcc ground state of Nb and the diffusionless transition paths to (b) fcc ($c/a=\sqrt{2}$, light blue) and bct (gray) as well as 
     (c)  Pnma (red) and (e) hexagonal $\omega$ (green)
     (d) and (f) show the unit cells of the relaxed Pnma and $\omega$ phases, respectively.}
    \label{fig:transitions}
\end{figure}
\subsection{Technical details}
Density-functional theory (DFT) simulations are performed with the abinit package.\cite{gonze_recent_2016}
All  calculations are done with primitive cells and Ta substitution is determined with the smallest possible cell up to a size of $2\times 2 \times 2$.
The results are determined with the generalized gradient approximation (GGA) using the Perdew-Burke-Ernzerhof (PBE) exchange-correlation functional~\cite{Perdew-96} in connection with optimized norm-conserving Vanderbilt pseudopotential (ONCVPSP) from PseudoDojo \cite{hamann_optimized_2013} with the valence electron con\-fi\-gu\-ra\-tion $4d^45s^1$ for Nb and $4f^{14} 5d^3 6s^2$ for Ta.
The stopping criterion for the self-consistent calculations is a difference in total energy of $2.72\cdot 10^{-7}$~eV. The volume and ionic positions of all structures are relaxed simultaneously with a tolerance on the maximal force of \SI{2.5e-3}{eV}/\AA\, and a smearing parameter for the total energy of \SI{95}{eV}, considering a temperature of smearing of \SI{0.272}{eV}, which is equivalent to \SI{3.15}{K}.
The $k$-mesh is $14\times 14 \times 14$ and a cutoff energy of \SI{1088.45}{eV} results in an energy convergence of \SI{0.172}{eV}.
\begin{table*}[tb]
    \centering
    \caption{List of labels, space groups, lattice parameters $a$, tetragonal ratios $c/a$ and energy differences to bcc of the (meta)-stable phases of interest. Available values from literature are added for comparison. Pnma$_\mathrm{max}$ and $\omega'$ refer to extrema from calculated transition paths (see Section~\ref{section: transformation paths}). 
    All values are based on PBE ($\dagger$: abinit,  $*$: VASP), except Refs. \onlinecite{mehl_applications_1996} and  \onlinecite{schoenecker_theoretical_2011} which are based on LDA. 
     If not noted in brackets 
   $a=b$. The tetragonal ratio of Pnma, $\omega$ and $\omega'$ in a pseudo-cubic unit cell are 1.73 , 0.83 and 0.89, respectively.}
     \begin{tabular}{c|c|c|c|c|c|l}
    %\hline
     \multirow{2}{*}{Label} & Space  & $c/a$ & $a$ & $V$ & $\Delta E$  & \multirow{2}{*}{Ref.}\\
    &group & ($b/a$) &(\AA)&(\AA$^3$/atom) &(meV/atom)\\
    \hline 
    \multirow{3}{*}{bcc} & \multirow{3}{*}{\textit{Im$\bar{3}$m}} & 1 & 3.30 & & 0.0 & \cite{nnolim_theoretical_2003}\\
    &&1& 3.31 & 18.11 &0.0&this work*\\
    && 1 &3.30 & 18.09 & 0.0 & this work$\dagger$ \\
    \hline
    \multirow{2}{*}{$\sigma$} &\multirow{2}{*}{\textit{{P$4_{\bar{2}}$}/mnm}}& 0.53 & 10.18 & 18.57 & 82 &this work*\\
    && 0.53 & 10.18&18.55 & 83 & this work$\dagger$\\
    \hline
   \multirow{3}{*}{A15}&\multirow{3}{*}{\textit{Pm$\bar{3}n$}}& 1 & 5.29 & & 104 &\cite{fellinger_force-matched_2010} \\
    &&1& 5.29& 18.56 &103&this work*\\
    && 1 & 5.29& 18.60 & 105 & this work$\dagger$ \\
    \hline
    \multirow{6}{*}{Pnma}&\multirow{6}{*}{\textit{Pnma}}&  0.88 & \multirow{2}{*}{5.11} & &\multirow{2}{*}{-\bibnote{Note that only a rough estimate of the energy has been reported (200~meV/atom) in Ref.~\onlinecite{errandonea_experimental_2020}. }} & \multirow{2}{*}{\cite{errandonea_experimental_2020}}\\
    && (0.52) & & &  & \\
    && 0.90 & \multirow{2}{*}{5.39} & \multirow{2}{*}{18.36} &\multirow{2}{*} {111} & \multirow{2}{*} {this work*} \\
    && (0.52) &  &  &  &  \\
    && 0.89 & \multirow{2}{*}{5.41} & \multirow{2}{*}{18.32} &\multirow{2}{*} {119} & \multirow{2}{*} {this work$\dagger$} \\
    && (0.51) &  &  &  & \\
    \hline
    \multirow{2}{*} {$\omega'$}& \multirow{2}{*}{\textit{P$\bar{3}$m$1$}} & 0.51 & 5.02&18.60 & 121 & this work$\dagger$\\
    && & & & {-\bibnote{Note that this phase is a minimum on the transition path, but not (meta)-stable}} & \\
    \hline
   \multirow{3}{*}{ bct}&\multirow{3}{*}{\textit{I$4$/mmm}}& 1.8 & 2.74 & & 180 & \cite{schoenecker_theoretical_2011}\\
    && 1.79 & 2.74 & 18.45 & 143 & this work*\\
    & & 1.77 & 2.71 & 18.48 & 143 & this work$\dagger$\\
    \hline
    \multirow{3}{*}{$\omega$}&\multirow{3}{*}{\textit{P$6$/mmm}} & 0.547 & 4.88 & & 201 & \cite{fellinger_force-matched_2010}\\
    && 0.548 & 4.88 & 18.45 & 199 & this work*\\    
    & & 0.550 & 4.88 & 18.47 & 202 & this work$\dagger$\\
    \hline
    \multirow{3}{*}{A13} &\multirow{3}{*}{\textit{P$4_132$}}& 1 & - & & 286 & \cite{mehl_applications_1996}\\
    && 1 & 7.21 & 18.69 & 210 &this work* \\
    && 1 & 7.14 & 18.77 & 223 &this work$\dagger$ \\
    \hline
    \multirow{2}{*}{Pnma$_{\mathrm{max}}$}& \multirow{2}{*}{\textit{Pnma}} & 0.67& \multirow{2}{*}{6.02} & \multirow{2}{*}{18.72} & \multirow{2}{*}{234} & \multirow{2}{*}{this work$\dagger$} \\
    &&(0.50)&&&&\\
    \hline
     \multirow{3}{*}{hcp} &\multirow{2}{*}{\textit{P$6_3$/mmc}} & 1.82 & 2.86 & & 297 & \cite{fellinger_force-matched_2010}\\
     & &1.83& 2.85 & 18.64 & 294 &this work*\\
    & & 1.82  & 2.87 & 18.66 & 294 & this work$\dagger$\\
    \hline
    \multirow{3}{*}{fcc} &\multirow{3}{*}{\textit{Fm$\bar{3}$m}} & 1 & 4.21 &  & 324 &\cite{fellinger_force-matched_2010}\\
    && 1 & 4.21 & 18.74 & 323 &this work*\\
    & & 1 & 4.21 & 18.77 & 323 & this work$\dagger$\\
    \end{tabular}
    \label{tab:lit_overview}
\end{table*}

Density-functional perturbation theory (DFPT) is used to determine phonon spectra and the phonon density of states.
As an insufficient sampling of q-space for bcc Nb results in the extrapolation to imaginary phonon modes,\cite{Souvatzis_ab-initio_2008} we chose a  grid of   $8\times 8 \times 8$ which is sufficient to correctly reproduce the experimental and theoretical data from literature.
Furthermore, we raise the threshold to a difference in potentials up to $10^{-18}$, the $k$-mesh to $16\times 16 \times 16$ and decreased the temperature of smearing to \SI{0.136}{eV} (corresponds to \SI{1.57}{K}). 
The $k$-mesh and q-grid for the other structures are scaled according to their lattice vectors. 
The phonon contribution to Helmholtz free energy $F_{\text{phon}}$ is calculated according to Lee et al.\cite{lee_ab_1995}
\begin{equation}\label{eq:free}
F_{\text{phon}}(T)=3nN k_BT\int_{0}^{\omega_L} \mathrm{ln}(2 \mathrm{sinh} \frac{\hbar\omega}{2k_BT})g(\omega)d\omega
\end{equation}
with the number of atoms per unit cell $n$, the number of unit cells $N$, the Boltzmann constant $k_B$ and temperature~$T$. $\omega_L$ is the largest frequency in the phonon spectra.
Without anharmonic effects and thermal expansion, the total free energy $F_{\text{total}}(T)$ is approximately given as 
\begin{equation}
    F_{\text{total}}(T)=E_{\text{tot}}(T=0~\text{K})+F_{\text{phon}}(T).
\end{equation}

For comparison we compute selected properties also with the VASP package~\cite{Kresse-96,Kresse-96b,Kresse-99} using the high-throughput environment from Ref.~\onlinecite{hammerschmidt_topologically_2013}. We use the PBE functional~\cite{Perdew-96} as in the abinit calculations but the projector-augmented wave method~\cite{Bloechl-94} and pseudo-potentials with $s$ semicore states for Nb and $p$ semicore states for Ta. With a planewave cut-off energy of \SI{500}{eV} and a $k$-point density of \SI{0.018}{\angstrom^3} we achieve similar convergence of the total energy differences as in our calculations with abinit.

Besides bcc, bct, $\omega$, Pnma, A13, A15, and $\sigma$ phases shown in Fig.~\ref{fig:fig3}, we also consider the Laves phases C14, C15, C36 and the structures $\mu$ and $\chi$. Ordered binary structures based on fcc or hcp are excluded due to the expected comparably high formation energy.~\cite{hammerschmidt_topologically_2013} For the considered phases, all occupations of Nb and Ta on the Wyckoff sites are included in the DFT calculations, e.g., $2^5=32$ DFT calculations for the $\sigma$ phase with five Wyckoff sites.
We asses the relative stability of the different structures and stoichiometries based on the heat of formation
\begin{equation}
    \Delta H_f=\frac{E_{\text{tot}} -N_{\text{Nb}}E_{\text{Nb}}-N_{\text{Ta}} E_{\text{Ta}}}{N}, 
\end{equation}
with $N_{\text{Nb}}$ ($N_{\text{Ta}}$) and $E_{\text{Nb}}$ ($E_{\text{Ta}}$) the number of Nb (Ta) atoms and their energies in the bcc phases. For pure systems this equation reduces to the energy difference of a structure to the bcc ground state, i.e., $\Delta E$ in Tab.~\ref{tab:lit_overview}.

\subsection{Transformation paths}\label{sec.transformPaths}

We study potential martensitic transition paths from bcc to fcc, $\omega$ and Pnma using a linear interpolation of the lattice constants $a_i$  to the final state as

\begin{equation}\label{eq:non-v-cons2}
   a_i(\Delta)=(1-\Delta) \cdot a_i^{\text{initial}}+\Delta \cdot a_i^{\text{final}}
\end{equation}
with varying $\Delta$ from $0$ to 1, see Fig.~\ref{fig:transitions}. 
The bcc to fcc transition is fully characterized by the lattice constants (see Fig.~\ref{fig:transitions}~(a)). For the bcc to Pnma and $\omega$ transition path, we additionally interpolate the internal atomic degrees of freedom linearly. 
The interpolation along the bcc-$\omega$ path (Fig.~\ref{fig:transitions}~(c)), corresponds to an anti-parallel shift of two of three atoms along z-direction from $\Delta_z=0$ for bcc to $\Delta_z=\pm 1/6$ for $\omega$ phase. Only in the $\omega$ phase, both these atoms are in the same z-plane of the hexagonal lattice (i.e. the same [111] plane of bcc) and the symmetry is P6/mmm while the symmetry is reduced to P$\bar{3}$m1 on the path.\cite{aurelio_interatomic_1999,garces_omega_1999} 

Note, that the transition from bcc to Pnma contains two unit cells of bcc, and $a_{\text{bcc}}$ consequently must be doubled, see Fig.~\ref{fig:transitions}~(b).
Two atoms shift by $\Delta_x$ and one atom each shift by $\Delta_x$ and $\Delta_{z1}$ and  another by $\Delta_{z2}$. For the relaxed Pnma structure we find values of $\Delta_{x}=0.01$, $\Delta_{z1}=-0.2$, $\Delta_{z2}=0.04$ , respectively. The atomic environment of the Pnma phase is thereby similar to that of the bcc state, but the 8-fold coordination (with a distance of \SI{2.86}{\angstrom}) splits into four nearest neighbours with average distances about \SI{2.80}{\angstrom} and four nearest neighbours with average distance of \SI{2.97}{\angstrom}.
Further we extrapolate the range of $\Delta$  to smaller or larger values to explore the energy landscape around the given states and analyse the A15 phase under tetragonal distortion in the range of $0.8<c/a<1.2$.

\section{Results}
\begin{figure}[t]
    \centering
    \includegraphics[width=0.48\textwidth,clip,trim=3mm 6mm 0cm 3mm]{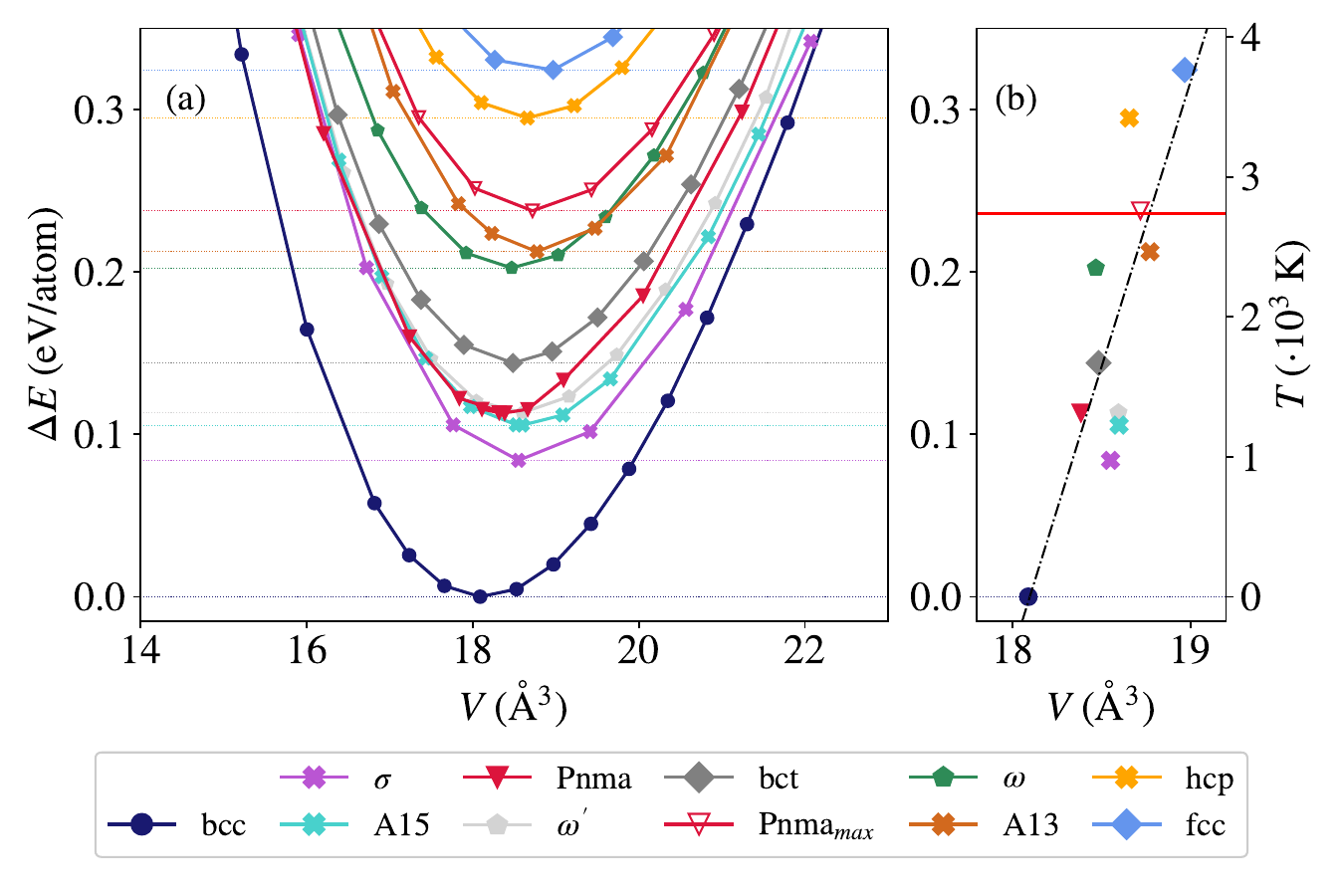}
    \caption{(a) Energy-volume curves of the (meta-)stable phases of Nb of interest. The energy per atom is given relative to the bcc ground state and (b) a reduced graphical representation with the minima for the selected structures only.
  The thick red horizontal line in (b) marks the melting temperature ($T_m=2741$~K) of Nb, \cite{cardarelli2008materials} according to the second axis with an estimation for $T$ using $E=k_\mathrm{B}T$. The black line represents a linear fit of the data.}
    \label{fig:fig3}
\end{figure}

\subsection{Comparison of meta-stable phases}
We consider all Nb structures listed in Tab.~\ref{tab:lit_overview}.
Figure~\ref{fig:fig3}~(a) shows their energy differences to the ground state, $\Delta E$, as a function of volume $V$ per atom. 
In agreement to literature we find bcc as the ground state structure while fcc is the least favourable of all tested configurations and about \SI{330}{meV} higher in energy. 
Still this phase has been observed in experiments underlining the importance of meta-stable Nb phases with lower energies.
Although the energy of the fcc phase is reduced by tetragonal distortion to the bct phase, A15 and Pnma structures are even more favourable with energy differences of about \SI{105}{meV} and \SI{119}{meV} relative to bcc only, respectively. The $\omega$, A13, and hcp phases are energetically between bct and fcc.

Our calculations show that the $\sigma$ phase is even more favourable being only slightly less than \SI{80}{meV} above the ground state. As we discuss in Sec.~\ref{section: transformation paths}, the bcc to $\omega$-phase shows an additional local minimum, which we have added as $\omega'$ for completeness. This configuration shows a minimum of the energy vs. volume curve only \SI{121}{meV} higher in energy than bcc. As discussed below the structure is however not stable against atomic relaxation. 

The same data is reduced to the minima of the E(V) curves in Fig.~\ref{fig:fig3}~(b). The energy of the metastable phases scales approximately linearly with the volume. Although, Pnma and $\omega$ have been predicted as high-pressure phases their ground state volumes without pressure is larger than the ground state volume of bcc.
The bcc structure is not only the energetic ground state for its equilibrium volume but also in the studied volume range of 15 to \SI{22}{\angstrom^3}.
Under lattice expansion between \SI{21}{\angstrom^3} and \SI{22}{\angstrom^3} the energy differences between bcc and $\sigma$ and A15 are however systematically reduced. Furthermore, in this volume range $\omega$ and A13 phases as well as Pnma and bct phases are close in energy.
On the other hand, none of the phases comes close to bcc for reduced volumes while the energy differences between $\omega'$, Pnma, A15 and $\sigma$ phases are reduced.

\begin{figure}[t]
    \centering{
    \includegraphics[width=0.48\textwidth,clip,trim=3mm 16mm 15mm 15mm]{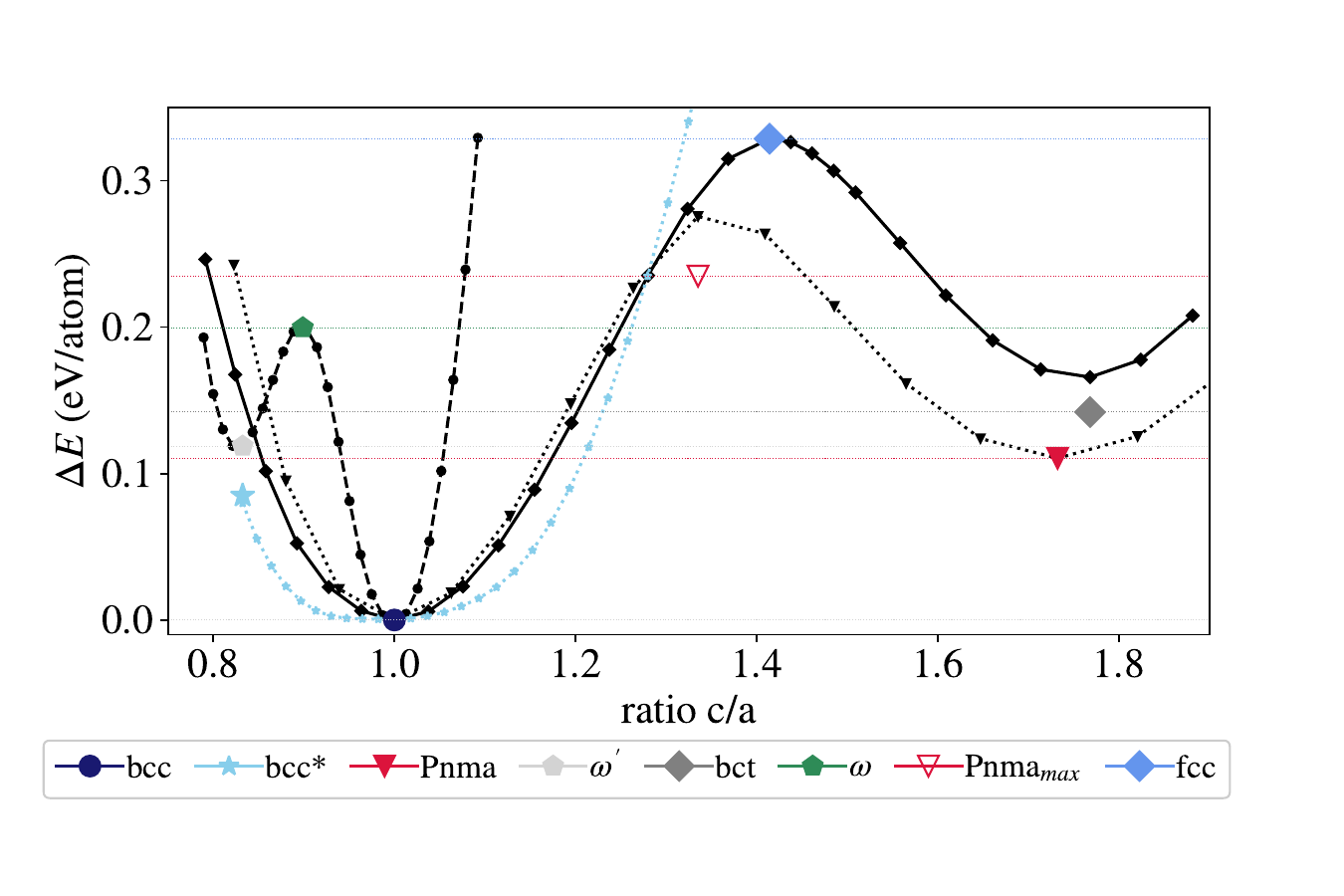}}
    \caption{Energy landscape for the deformation of the bcc cubic structure with $c/a=1$ along (a) the Bain path to fcc (solid line with diamonds), (b) the path to Pnma (dotted line with triangles) and (c) the path to $\omega$ (dashed line with circles). The lattice parameters of Pnma and $\omega$ are given in a pseudo-cubic unit cell. All colored symbols indicate structures whose volumes and positions of atoms are relaxed, unlike the other data points. For bcc in the hexagonal $\omega$ cell (bcc*), furthermore the energy landscape for tetragonal distortion is shown in light blue.
}
    \label{fig16}
\end{figure}

\subsection{Transformation paths} \label{section: transformation paths}
To depict possible diffusionless phase transitions of Nb, we study the 
energy landscape for a continuous deformation from bcc to meta-stable structures. The energy maxima on theses paths give an upper bound for the energy barriers of the transitions. Note, that the real energy barriers can be smaller due to their dependence on temperature or more complex transition paths.
As reference we start with the classical Bain path from bcc to fcc (solid lines in Fig.~\ref{fig16}). In agreement with literature  we find that fcc is a local maximum on the transformation path and the second minimum, the bct structure with $c/a=1.768$  is \SI{143}{meV} higher in energy than bcc, see Tab.~\ref{tab:lit_overview}. 

In the following we restrict ourselves to the low energy structures shown in Fig~\ref{fig:transitions}: (a) bcc, (d) Pnma and (f) $\omega$.
The dotted lines in Fig.~\ref{fig16} show the bcc to Pnma path. 
On this path, only the two energy minima related to bcc and Pnma occur and both states are separated by an energy barrier of \SI{234}{meV} which is about \SI{89}{meV} lower than the fcc state.
Surprisingly, the $\omega$ phase is not even a local minimum of energy on the bcc to $\omega$ path but rather a local energy maximum. By extrapolation of $\Delta$ we find an energy minimum ($\omega'$) with $\Delta z=0.28$ and $c/a=0.832$ only \SI{121}{meV} higher in energy than the bcc ground state. Note that this anomaly has not been found for a variation of $\Delta z$ with fixed tetragonal ratio\cite{garces_omega_1999} and the structure differs from the modulated $\omega$ structure with vacancies discussed in the supplementary material from Ref.~\onlinecite{lee_stress-induced_2022}. 
Although this monoclinic configuration also shows the typical energy volume curve of a meta-stable state, see Fig.~\ref{fig:fig3}, the atomic positions are not protected against atomic relaxation by symmetry. Only the $\omega$ phase with P6/mmm symmetry is a meta-stable state, while the atomic positions relax to the bcc-like structure with $z=0$ for all other initial values of $z$.

The distorted bcc state at $c/a=0.832$, which we call bcc* in the following, is only \SI{85}{meV} higher in energy than bcc and may thus be a favorable distortion of the bcc phase.
For this reason we sample the energy for the tetragonal distortion of the bcc* 
to the bcc state as shown in Fig.~\ref{fig:fig3}. 
Our calculations show that the energy penalty for the distortion along the [111] direction is considerably smaller than for the bcc to fcc path.

Given the lack of information on transition paths among the metastable states, we additionally verified if tetragonally distorted structures are  possible for other metastable phases of Nb. Particularly the cubic A15 phase is low in energy and we also study its tetragonal distortion, see Fig.~\ref{fig:fig3}. But even with a fine resolution of $\Delta c/a=2\cdot 10^{-4}$ we could not observe any additional local minima or higher-order extrema under tetragonal distortion. The increase of energy with tetragonal distortion is similar to the classical Bain path.

\subsection{Phonon spectra}
\begin{figure}[t]
    \centering
    \includegraphics[width=0.49\textwidth, clip,trim=4mm 0mm 18.5mm 6.5mm]{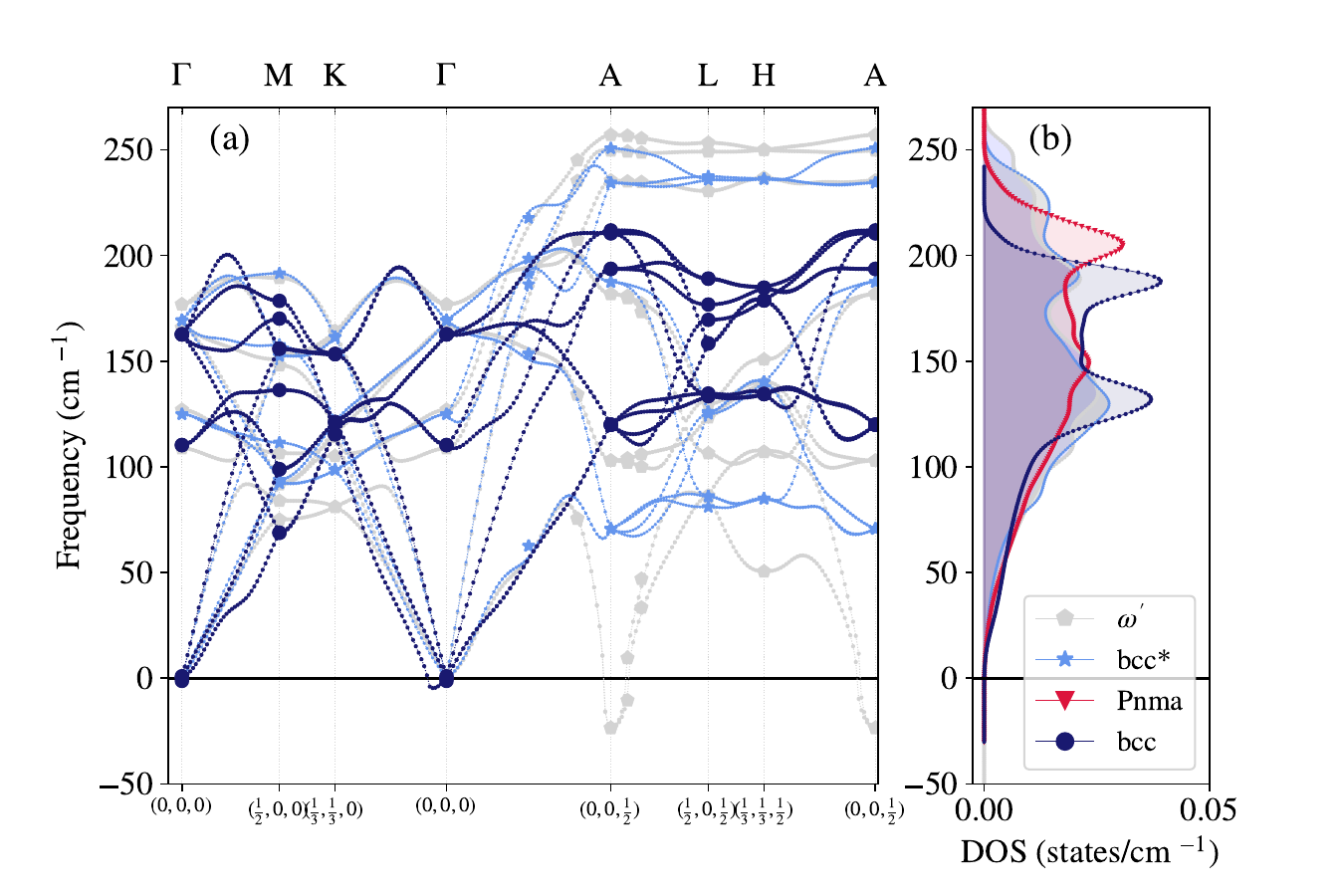}\\
    \caption{(a) Phonon spectra of the $\omega'$ phase (grey) bcc* (light blue) and bcc (dark blue). The dots mark q-points used in the DFPT framework and lines are interpolations. (b) Comparison of the phonon densities of state (DOS) per atom for $\omega'$, bcc*, Pnma (red) and bcc. 
    %The phonon spectra of Pnma and bcc in a primitive cell can be found in the appendix. 
    }
    \label{fig:phonon}
\end{figure}

\begin{figure}[t]
    \centering
    \includegraphics[width=0.49\textwidth, clip,trim=4mm 0mm 18.5mm 6.5mm]{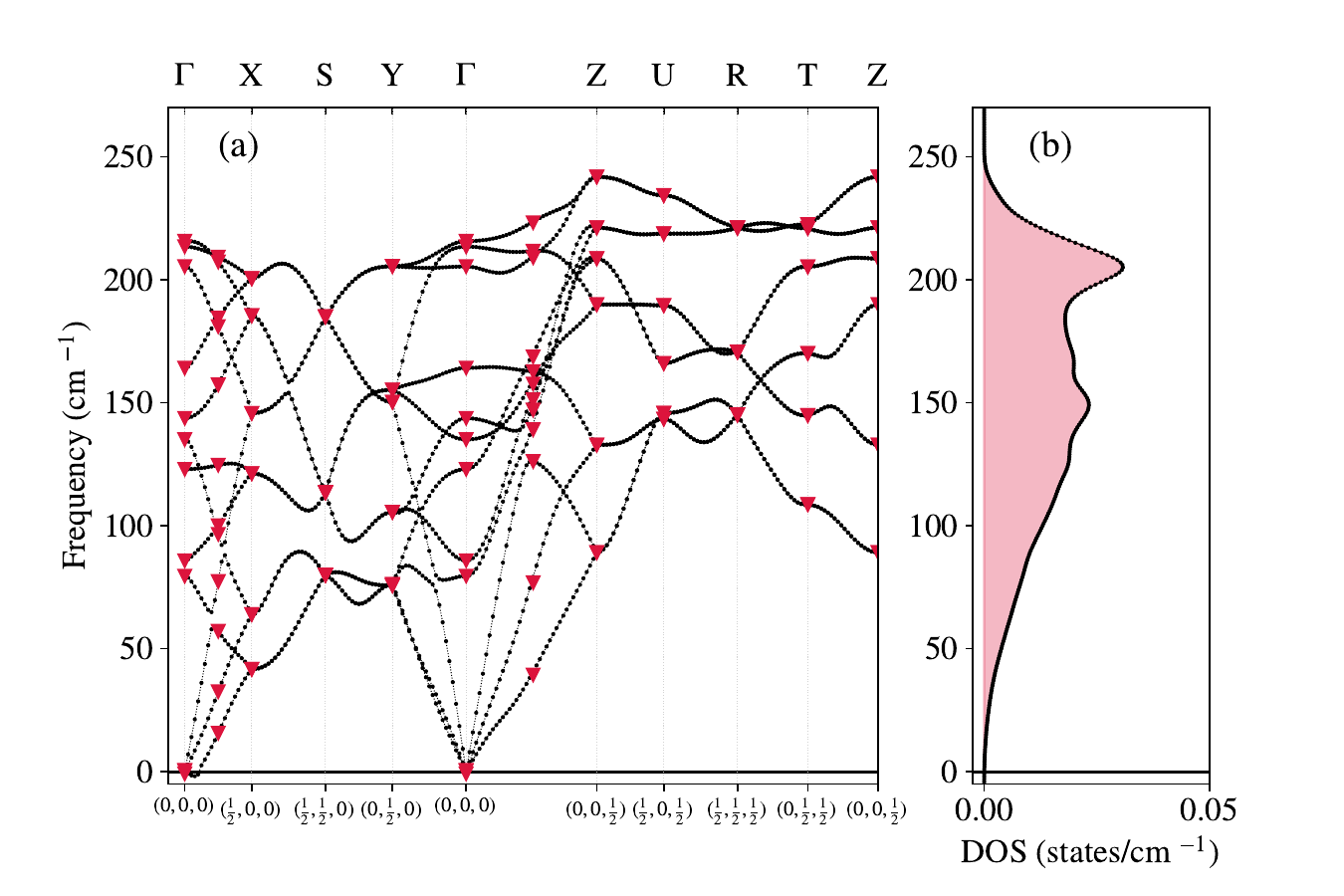}\\
    \caption{(a) Phonon spectrum of the Pnma phase. The dots mark q-points used in the DFPT framework and lines are interpolations. (b) The phonon densities of state (DOS) per atom.
    }
    \label{fig:phonon_pnma}
\end{figure}

For a comprehensive picture of the low-energy phases, we analyze the phonon spectra of bcc, bcc* (bcc phase distorted along [111] with $c/a=0.832$), $\omega'$ and Pnma.
While the phonon spectra of the bcc and Pnma structures calculated by us correspond to those published in literature,\cite{marzi_electronic_2016, errandonea_experimental_2020}, we are not aware of calculated phonon spectra of (distorted) $\omega$ phases which are summarized in Fig.~\ref{fig:phonon}~(a). As reference we also added the phonon spectra of bcc in the same representation. 

Indeed, the ground state, bcc, only shows stable phonon modes. However, there are indications, that the structure is close to an instability. Particularly, we can reproduce the Kohn anomaly at $(0.142,-0.142,0.142)$ predicted by Landa et al.\cite{landa_kohn_2018} Note, that in the representation in Fig.~\ref{fig:phonon}~(a) this point is located on the $\Gamma \rightarrow$ M path.
Furthermore we can reproduce the decrease of the transversal acoustic branch in the phonon spectra  at $(1/3,0,1/3)$ in the $[111]$ direction associated with the bcc to $\omega$ transition.\cite{cook_theory_1974,garces_omega_1999}

The metastable Pnma phase also shows no soft phonons, see Fig.~\ref{fig:phonon_pnma}~(a). Compared to bcc the change in slope on the $\Gamma \rightarrow$ Z path is reduced. Furthermore, due to the lower symmetry of Pnma one has to distinguish X, Y and Z direction and for the former two we see no change in slope on the corresponding paths with the given resolution.
Analogous to bcc and Pnma, also bcc* does not show soft phonon modes.
Under the hexagonal distortion the change of the slope on the $\Gamma \rightarrow$ M path vanishes, both for bcc* and $\omega'$, while the lowest $\Gamma \rightarrow$ A branch shows a similar feature.
Moreover, for all high-symmetry points except M, the transversal branches are lowered in energy if going from bcc to  $\omega'$ and bcc*.
As discussed in Sec.~\ref{sec.transformPaths} the $\omega'$ structure, although being a local energy minimum on the bcc--$\omega$ path is not a stable structure and thus the phonon spectrum shows negative frequencies at A (0,0,1/2).  

Figure~\ref{fig:phonon}~(b) compares the resulting phonon density of states of all four structures normalized with the number of atoms in the system.
Over a large frequency range from \SI{125}{cm^{-1}} to \SI{200}{cm^{-1}}, the bcc phase shows the largest density of states with two pronounced peaks around \SI{135}{cm^{-1}} and \SI{190}{cm^{-1}}. 
With decreasing symmetry going from bcc to bcc* and $\omega'$ the degeneracies of the modes in [100] direction are lifted and the peaks in the DOS are broadened. The distortion of the structure to $\omega'$ furthermore results in two additional peaks at \SI{80}{cm^{-1}} and \SI{230}{cm^{-1}} and a low-frequency tail.
Below frequencies of \SI{125}{cm^{-1}}, the $\omega'$ phase thus exhibits the highest density of states.
A lifting of degeneracies in the modes in [100] direction can also be seen for the Pnma phase. Also here, we find higher frequencies in the spectrum and less pronounced peaks in the DOS.
For the Pnma phase the largest DOS is found at about \SI{210}{cm^{-1}} and the increase of the weight of the low-frequency tail of the DOS is slightly larger than for $\omega'$.

However, none of the phases has a substantially larger phonon DOS in a suitable frequency range and the low-frequency tails of the DOS are not sufficient to reduce the free energy and to stabilize one of the phases relative to bcc at finite temperatures. Note that we calculated the free energies 
within the harmonic approximation and thus higher order effects are not accounted in this estimate.

\subsection{Ta-substitution}
\begin{figure}[t]
    \includegraphics[trim = 2.5mm 15mm 10mm 30mm, clip, width=0.48\textwidth]{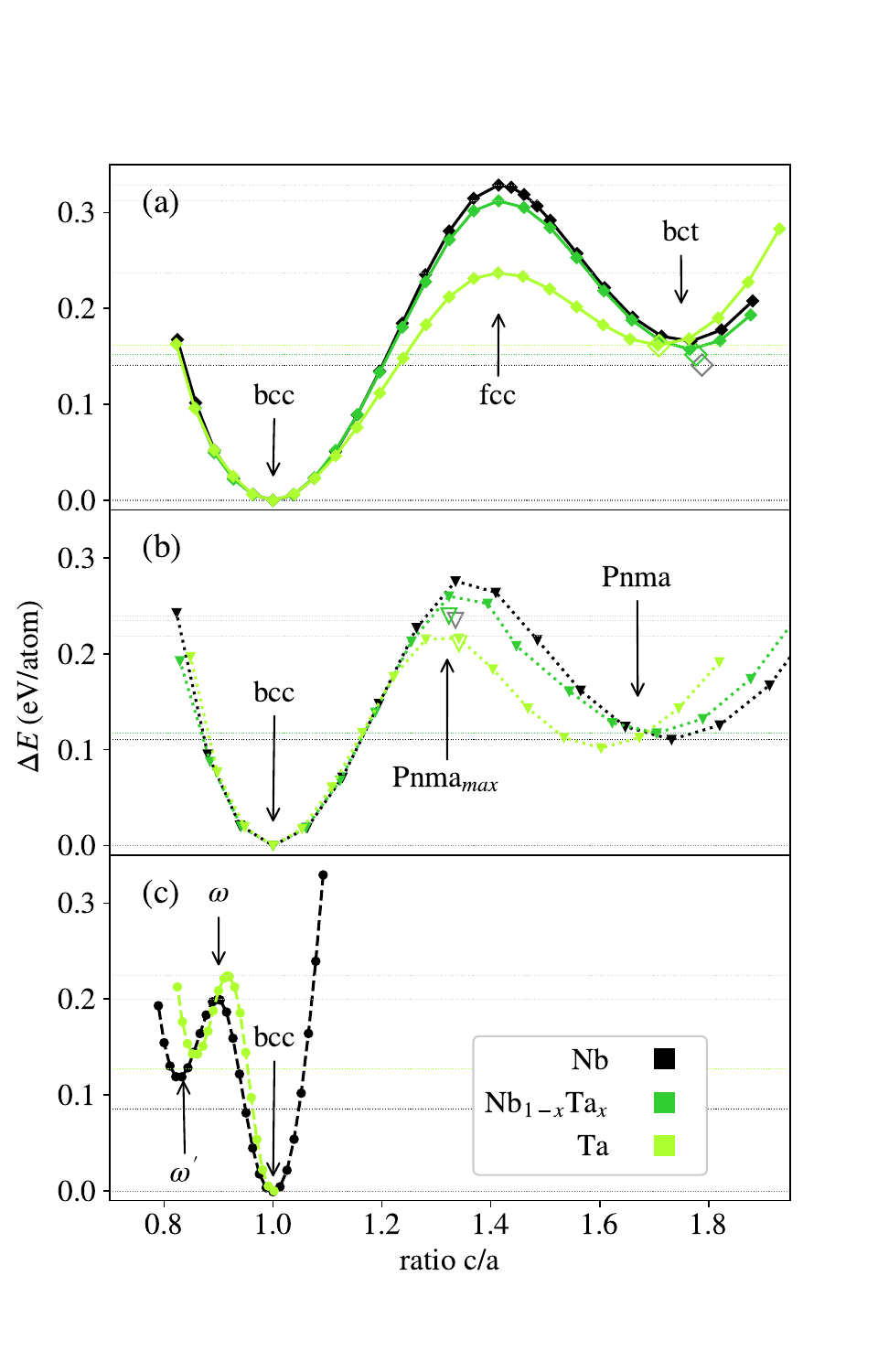}
 \caption{Impact of Ta-substitution with a concentration of $x=0.25$ on (a) the bcc-bct Bain transition path, (b) the bcc to Pnma transition path with two bcc unit cells and a random Ta substitution and (c) the bcc to $\omega$ transition path. Note that volume and atomic positions have been fixed on the transition path (filled symbols) and open symbols indicate the energy of the energy maximum after corresponding relaxation.
 }
    \label{pnmatobcc2}   
\end{figure}
In order to enable a close comparison of our results with experimental works, we additionally consider the influence of Ta on the metastable phases of Nb, the most common impurity in Nb samples for experiments.

Figure~\ref{pnmatobcc2} compares the energy landscapes of pure Nb (black) and Ta (light green) along the classical Bain path (a) and  the paths connecting bcc (b)  and Pnma  (c) and $\omega$ phases.
In all cases, the energetic ground state is the bcc state with $c/a=1$ and the other local energy minima are not considerably lowered by Ta. 
The ratio of the lattice constants c/a is smaller for Ta compared to Nb ($-0.07$ for bct, and $-0.13$ for Pnma), but slightly increased for $\omega'$ by $0.02$. 
For bct (panel a) and Pnma (panel b), the energy barrier for the transformation is \SI{91}{meV} and \SI{24}{meV} smaller for Ta compared to Nb. 
However, the changes of the energy landscapes under partial substitution are small. Exemplary results for 25\% Ta are added in panels (a) and (b) in dark green. 
Even for this large concentration, the energy differences between the pure and substituted materials are below \SI{16}{meV} and \SI{4}{meV} at the transition barrier. 
For the bcc to $\omega$ transition, the barrier is not smaller for Ta compared to Nb and for both elements the structure at the second minimum is not stable against atomic relaxation. 
\begin{figure}[t]
    \includegraphics[clip,trim=0cm 0cm 0cm .2cm, width=0.48\textwidth]{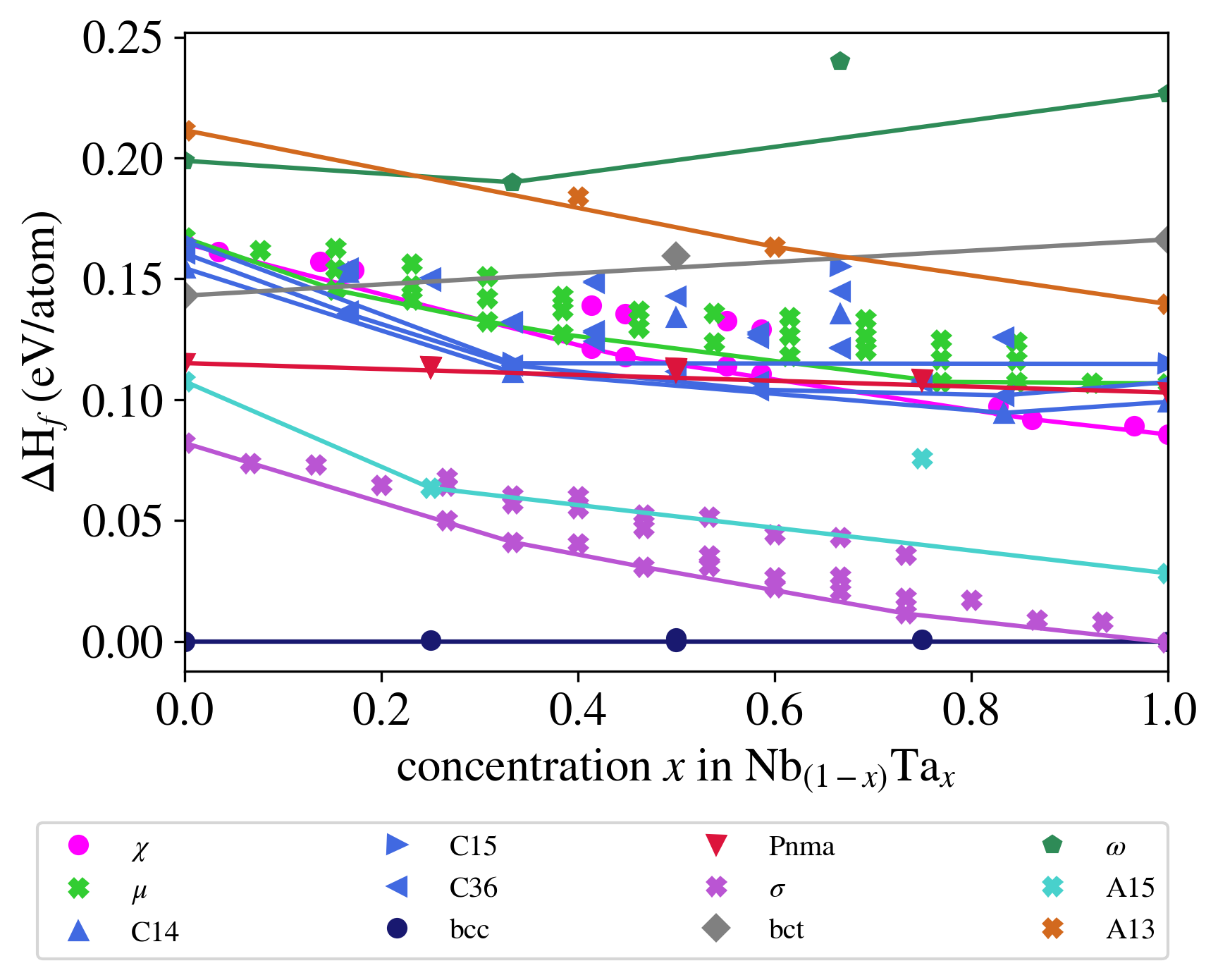}
    \caption{Structural stability of Nb-Ta phases obtained by DFT with permutation of Nb and Ta atoms on Wyckoff sites of bcc, bct, C32, A13 and the topological complex phases (TCP) phases A15, C14, C15, C36, $\mu$, $\chi$ and $\sigma$. The lines represent the convex hulls of the individual phases to guide the eye.}
    \label{fig:NbTa}   
\end{figure}

The influence of Ta on the structural stability of all phases of interest across the complete range of chemical compositions is shown in Fig.~\ref{fig:NbTa} in terms of the formation energies $\Delta H_f$.
The relative stability for pure Nb is identical to the sequence of minima observed in the energy-volume curves in Fig.~\ref{fig:fig3}. 
The variation across the Nb-Ta chemical range is consistent with previous DFT calculations~\cite{hammerschmidt_topologically_2013} using an LDA exchange-correlation functional although PBE shows slightly lower formation energies $\Delta H_f$. 
Comparing pure Ta to Nb, the phase sequence bcc, $\sigma$, A15, Pnma, bct and $\omega$ is still present. However, the A13, $\sigma$ and A15 phases are considerably lowered in energy. The $\sigma$ phase is very close to bcc, in line with the experimental characterization of $\beta$-Ta as $\sigma$ phase.~\cite{Jiang-03} 
The formation energies of the Pnma phase depend only  weakly on the Ta concentration.
For completeness also the formation energies of the Laves (C14, C15, C36), $\chi$ and $\mu$ phases are shown in Fig.~\ref{fig:NbTa}. 
For Nb all these phases are higher in energy than bct. There is no sizeable stabilisation of the Laves  and $\mu$ phases by Ta, while the $\chi$ phase becomes more favourable than the Pnma phase.
The throughout positive values of $\Delta H_f$ indicate that there is no stable ordered structure, in line with the bcc solid-solution region in the phase diagram.

Thus Ta indeed reduces the energy barrier for the bcc to bct or Pnma transitions but quantitatively the effect is small.
We would furthermore expect that alloying Nb with small amounts of Ta may foster the formation of A13 or $\sigma$ phases while an enhanced formation of Pnma is unlikely.
\\
\\
\section{Summary and Conclusions}
The question of potential metastable phases in Nb was raised anew by high-resolution experimental data suggesting a martensitic phase transition.\cite{bollinger_observation_2011}
To better understand the energy landscape of Nb, we determined the ground states for chosen metastable phases using DFT and analysed possible transition paths connecting these with the bcc ground state.
We find that the metastable $\sigma$ and A15 phases are lowest in energy  followed by Pnma, bct and A13. Both the bcc to Pnma and bcc to $\omega$ path are more favourable than the more commonly discussed Bain path to fcc. Additionally, straining bcc along the hexagonal [111] direction, we find a potential deformed state $\omega'$ being low in energy.

Since Ta impurities are common in Nb, we also investigated the role of Ta on the energy landscape of the metastable phases. The energy barriers for the bcc to bct and Pnma transition are reduced. In pure Ta, the $\sigma$ phase is practically as low in energy as the bcc phase. Thus, for high Ta concentrations, this is important and should be further investigated in the future.

Otherwise, considering our DFT study, we suggest that effects that could come from the microstructure not detected here - like stabilization of a metastable phase by twinning - may trigger phase transitions or modify the atomic structure in large parts of an experimental sample, explaining the experimentally found martensitic phase transition.

\begin{acknowledgments}
We thank Anna B\"ohmer and Ralf Drautz for fruitful discussion.
\end{acknowledgments}

\bibliographystyle{unsrt}
\bibliography{Nb}

\begin{thebibliography}{10}

\bibitem{sauls_2023_effects}
J.~A. Sauls, M.~Zarea, and H.~Ueki.
\newblock Effects of {A}nisotropy and {D}isorder on the {S}uperconducting {P}roperties of {N}iobium.
\newblock {\em Frontiers in Physics}, 11:1269872, 2023.

\bibitem{SELLERS}
G.J. Sellers, A.C. Anderson, and H.K. Birnbaum.
\newblock The anomalous heat capacity of superconducting niobium.
\newblock {\em Physics Letters A}, 44(3):173, 1973.

\bibitem{struzhkin_superconducting_1997}
V.~Struzhkin, Y.~Timofeev, R.~Hemley, and H.~Mao.
\newblock {S}uperconducting ${T}_c$ and {E}lectron-{P}honon {C}oupling in {Nb} to 132~{GPa}: {M}agnetic {S}usceptibility at {M}egabar {P}ressures.
\newblock {\em Physical Review Letters}, 79(21):4262, 1997.

\bibitem{li_shear_2021}
X.~Li, Q.~Zhao, Q.~Wang, Y.~Tian, H.~Zhou, and J.~Wang.
\newblock {S}hear band mediated $\omega$ phase transformation in {Nb} single crystals deformed at 77 {K}.
\newblock {\em Materials Research Letters}, 9(12):523, 2021.

\bibitem{errandonea_experimental_2020}
D.~Errandonea, L.~Burakovsky, D.~L. Preston, S.~G. {MacLeod}, D.~Santamaría-Perez, S.~Chen, H.~Cynn, S.~I. Simak, M.~I. {McMahon}, J.~E. Proctor, and M.~Mezouar.
\newblock {E}xperimental and theoretical confirmation of an orthorhombic phase transition in niobium at high pressure and temperature.
\newblock {\em Communications Materials}, 1(1):60, 2020.

\bibitem{wang_consecutive_2018}
Q.~Wang, J.~Wang, J.~Li, Z.~Zhang, and S.~X. Mao.
\newblock {C}onsecutive crystallographic reorientations and superplasticity in body-centered cubic niobium nanowires.
\newblock {\em Science Advances}, 4(7):eaas8850, 2018.

\bibitem{chattopadhyay_polymorphic_2001}
P.~P. Chattopadhyay, P.~M.~G. Nambissan, S.~K. Pabi, and I.~Manna.
\newblock {P}olymorphic bcc to fcc transformation of nanocrystalline niobium studied by positron annihilation.
\newblock {\em Physical Review B}, 63(5):054107, 2001.

\bibitem{bollinger_observation_2011}
R.~K. Bollinger, B.~D. White, J.~J. Neumeier, H.~R.~Z. Sandim, Y.~Suzuki, C.~A.~M. dos Santos, R.~Avci, A.~Migliori, and J.~B. Betts.
\newblock {O}bservation of a {M}artensitic {S}tructural {D}istortion in {V}, {Nb}, and {Ta}.
\newblock {\em Physical Review Letters}, 107(7):075503, 2011.

\bibitem{grunebohm_unifying_2023}
A.~Gr{\"u}nebohm, A.~H{\"u}tten, A.~E. B{\"o}hmer, J.~Frenzel, I.~Eremin, R.~Drautz, I.~Ennen, L.~Caron, T.~Kuschel, F.~Lechermann, D.~Anselmetti, T.~Dahm, F.~Weber, K.~Rossnagel, and G.~Schierning.
\newblock A {{Unifying Perspective}} of {{Common Motifs That Occur}} across {{Disparate Classes}} of {{Materials Harboring Displacive Phase Transitions}}.
\newblock {\em Advanced Energy Materials}, 13(30):2300754, 2023.

\bibitem{nakagawa_lattice_1963}
Y.~Nakagawa and A.~D.~B. Woods.
\newblock {Lattice Dynamics} of {Niobium}.
\newblock {\em Physical Review Letters}, 11(6):271, 1963.

\bibitem{de_gironcoli_lattice_1995}
S.~de~Gironcoli.
\newblock Lattice dynamics of metals from density-functional perturbation theory.
\newblock {\em Physical Review B}, 51(10):6773, 1995.

\bibitem{landa_kohn_2018}
A.~Landa, P.~Söderlind, I.~Naumov, J.~Klepeis, and L.~Vitos.
\newblock {Kohn Anomaly} and {Phase Stability} in {Group {VB} Transition Metals}.
\newblock {\em Computation}, 6(2):29, 2018.

\bibitem{tidholm_temperature_2020}
J.~Tidholm, O.~Hellman, N.~Shulumba, S.~I. Simak, F.~Tasnádi, and I.~A. Abrikosov.
\newblock Temperature dependence of the {K}ohn anomaly in bcc {Nb} from first-principles self-consistent phonon calculations.
\newblock {\em Physical Review B}, 101(11):115119, 2020.

\bibitem{Aynajian_2008}
P.~Aynajian, T.~Keller, L.~Boeri, S.~M. Shapiro, K.~Habicht, and B.~Keimer.
\newblock Energy gaps and {K}ohn anomalies in elemental superconductors.
\newblock {\em Science}, 319(5869):1509, 2008.

\bibitem{liu_first_2011}
Zenghui Liu and Jiaxiang Shang.
\newblock First principles calculations of electronic properties and mechanical properties of bcc molybdenum and niobium.
\newblock {\em Rare Metals}, 30:354, 2011.

\bibitem{mehl_applications_1996}
M.~J. Mehl and D.~A. Papaconstantopoulos.
\newblock Applications of a tight-binding total-energy method for transition and noble metals: {E}lastic constants, vacancies, and surfaces of monatomic metals.
\newblock {\em Physical Review B}, 54(7):4519, 1996.

\bibitem{fellinger_force-matched_2010}
M.~R. Fellinger, H.~Park, and J.~W. Wilkins.
\newblock Force-matched embedded-atom method potential for niobium.
\newblock {\em Physical Review B}, 81(14):144119, 2010.

\bibitem{hammerschmidt_topologically_2013}
T.~Hammerschmidt, A.~F. Bialon, D.~G. Pettifor, and R.~Drautz.
\newblock Topologically close-packed phases in binary transition-metal compounds: matching high-throughput \textit{ab initio} calculations to an empirical structure map.
\newblock {\em New Journal of Physics}, 15(11):115016, 2013.

\bibitem{sasaki_fcc_1988}
M.~Sasaki, M.~Koyano, H.~Negishi, and M.~Inoue.
\newblock F.c.c. niobium films grown by halide chemical vapour deposition on ultrasound-vibrating substrates.
\newblock {\em Thin Solid Films}, 158(1):123, 1988.

\bibitem{lee_stress-induced_2022}
J.~Lee, Z.~Sung, A.~A. Murthy, A.~Grassellino, A.~Romanenko, N.~S. Sitaraman, and T.~A. Arias.
\newblock Stress-induced structural changes in superconducting {Nb} thin films.
\newblock {\em Physical Review Materials}, 7:L063201, 2023.

\bibitem{godeke_topical_2006}
A.~Godeke.
\newblock A review of the properties of {Nb$_3$Sn} and their variation with {A15} composition, morphology and strain state.
\newblock {\em Superconductor Science \& Technology}, 19, 2006.

\bibitem{Sadigh_1998_structural}
B.~{Sadigh} and V.~{Ozoli{\c{n}}{\v{s}}}.
\newblock {Structural instability and electronic excitations in Nb$_{3}$Sn}.
\newblock {\em Physical Review B}, 57(5):2793, 1998.

\bibitem{schoenecker_theoretical_2011}
S.~Schoenecker.
\newblock {\em Theoretical studies of epitaxial {B}ain paths of metals}.
\newblock dissertation, Technische Universität Dresden, 2011.

\bibitem{nnolim_theoretical_2003}
N.~O. Nnolim, T.~A. Tyson, and L.~Axe.
\newblock A theoretical study of the structural phases of {G}roup {5B – 6B} metals and their transport properties.
\newblock {\em Journal of Applied Physics}, 93:4543, 2003.

\bibitem{craievich_local_1994}
P.~J. Craievich, M.~Weinert, J.~M. Sanchez, and R.~E. Watson.
\newblock Local stability of nonequilibrium phases.
\newblock {\em Physical Review Letters}, 72(19):3076, 1994.

\bibitem{craievich_structural_1997}
P.~J. Craievich, J.~M. Sanchez, R.~E. Watson, and M.~Weinert.
\newblock Structural instabilities of excited phases.
\newblock {\em Physical Review B}, 55(2):787, 1997.

\bibitem{natarajan_connecting_2018}
A.~R. Natarajan and A.~Van~der Ven.
\newblock Connecting the {Simpler Structures} to {Topologically Close-Packed Phases}.
\newblock {\em Physical Review Letters}, 121(25):255701, 2018.

\bibitem{KOLLI}
S.~K. Kolli, A.~R. Natarajan, and A.~{Van der Ven}.
\newblock Six new transformation pathways connecting simple crystal structures and common intermetallic crystal structures.
\newblock {\em Acta Materialia}, 221:117429, 2021.

\bibitem{xiao_solid-state_2014}
P.~Xiao, D.~Sheppard, J.~Rogal, and G.~Henkelman.
\newblock Solid-state dimer method for calculating solid-solid phase transitions.
\newblock {\em The Journal of Chemical Physics}, 140(17):174104, 2014.

\bibitem{NETE}
M.~Nete, W.~Purcell, and J.T. Nel.
\newblock Separation and isolation of tantalum and niobium from tantalite using solvent extraction and ion exchange.
\newblock {\em Hydrometallurgy}, 149:31, 2014.

\bibitem{yao_stable_2013}
Y.~Yao and D.~D. Klug.
\newblock Stable structures of tantalum at high temperature and high pressure.
\newblock {\em Physical Review B}, 88(5):054102, 2013.

\bibitem{gonze_recent_2016}
X.~Gonze, F.~Jollet, F.~Abreu~Araujo, D.~Adams, B.~Amadon, T.~Applencourt, C.~Audouze, J.-M. Beuken, J.~Bieder, A.~Bokhanchuk, E.~Bousquet, F.~Bruneval, D.~Caliste, M.~Côté, F.~Dahm, F.~Da~Pieve, M.~Delaveau, M.~Di~Gennaro, B.~Dorado, C.~Espejo, G.~Geneste, L.~Genovese, A.~Gerossier, M.~Giantomassi, Y.~Gillet, D.R. Hamann, L.~He, G.~Jomard, J.~Laflamme~Janssen, S.~Le~Roux, A.~Levitt, A.~Lherbier, F.~Liu, I.~Lukačević, A.~Martin, C.~Martins, M.J.T. Oliveira, S.~Poncé, Y.~Pouillon, T.~Rangel, G.-M. Rignanese, A.H. Romero, B.~Rousseau, O.~Rubel, A.A. Shukri, M.~Stankovski, M.~Torrent, M.J. Van~Setten, B.~Van~Troeye, M.J. Verstraete, D.~Waroquiers, J.~Wiktor, B.~Xu, A.~Zhou, and J.W. Zwanziger.
\newblock Recent developments in the {ABINIT} software package.
\newblock {\em Computer Physics Communications}, 205:106, 2016.

\bibitem{Perdew-96}
J.~P. Perdew, K.~Burke, and M.~Ernzerhof.
\newblock {Generalized Gradient Approximation Made Simple}.
\newblock {\em Physical Review Letters}, 77:3865, 1996.

\bibitem{hamann_optimized_2013}
D.~R. Hamann.
\newblock Optimized norm-conserving {Vanderbilt} pseudopotentials.
\newblock {\em Physical Review B}, 88(8):085117, 2013.

\bibitem{Note1}
Note that only a rough estimate of the energy has been reported (200~meV/atom) in Ref.~\onlinecite {errandonea_experimental_2020}.

\bibitem{Note2}
Note that this phase is a minimum on the transition path, but not (meta)-stable.

\bibitem{Souvatzis_ab-initio_2008}
P.~Souvatzis and O.~Eriksson.
\newblock Ab initio calculations of the phonon spectra and the thermal expansion coefficients of the $4d$ metals.
\newblock {\em Physical Review B}, 77:024110, 2008.

\bibitem{lee_ab_1995}
C.~Lee and X.~Gonze.
\newblock \textit{Ab initio} calculation of the thermodynamic properties and atomic temperature factors of {SiO}$_2$ $\alpha$-quartz and stishovite.
\newblock {\em Physical Review B}, 51(13):8610, 1995.

\bibitem{Kresse-96}
G.~Kresse and J.~Furthm{\"u}ller.
\newblock Efficiency of ab-initio total energy calculations for metals and semiconductors using a plane-wave basis set.
\newblock {\em Computational Materials Science}, 6(1):15, 1996.

\bibitem{Kresse-96b}
G.~Kresse and J.~Furthm\"uller.
\newblock Efficient iterative schemes for ab initio total-energy calculations using a plane-wave basis set.
\newblock {\em Physical Review B}, 54:11169, 1996.

\bibitem{Kresse-99}
G.~Kresse and D.~Joubert.
\newblock From ultrasoft pseudopotentials to the projector augmented-wave method.
\newblock {\em Physical Review B}, 59:1758, 1999.

\bibitem{Bloechl-94}
P.~E. Bl\"ochl.
\newblock Projector augmented-wave method.
\newblock {\em Physical Review B}, 50:17953, 1994.

\bibitem{aurelio_interatomic_1999}
G.~Aurelio and A.~Fernández~Guillermet.
\newblock Interatomic distances in the stable and metastable bcc and omega structures of the transition metals: analysis of experimental and theoretical trends and correlations with {P}auling’s bond lengths.
\newblock {\em Journal of Alloys and Compounds}, 292(1):31, 1999.

\bibitem{garces_omega_1999}
J.E. Garcés, G.B. Grad, A.~{Fernández Guillermet}, and S.J. Sferco.
\newblock Theoretical study of the structural properties and thermodynamic stability of the omega phase in the 4d-transition series.
\newblock {\em Journal of Alloys and Compounds}, 289(1):1, 1999.

\bibitem{cardarelli2008materials}
Fran{\c{c}}ois Cardarelli.
\newblock {\em Materials handbook: a concise desktop reference}.
\newblock Springer London, 2008.

\bibitem{marzi_electronic_2016}
G.~D. Marzi.
\newblock {Electronic Band Structure, Lattice Dynamics, and Related Superconducting Properties of Niobium from First-Principles Calculations}.
\newblock Technical report, {ENEA}, 2016.

\bibitem{cook_theory_1974}
H.E. Cook.
\newblock A theory of the omega transformation.
\newblock {\em Acta Metallurgica}, 22(2):239, 1974.

\bibitem{Jiang-03}
A.~Jiang, A.~Yohannan, N.O. Nnolim, T.A. Tyson, L.~Axe, S.L. Lee, and P.~Cotec.
\newblock Investigation of the structure of $\beta$-tantalum.
\newblock {\em Thin Solid Films}, 437:116, 2003.

\end{thebibliography}

\end{document}